\documentclass[showpacs,nofootinbib,floats,floatfix,preprintnumbers,
groupedaddress,twocolumn]{revtex4}

\usepackage{bm}
\usepackage{latexsym}
\usepackage{dcolumn}
\usepackage{amsmath,amsfonts,amssymb, amsthm}
\usepackage{graphicx,epsfig}
\usepackage{environ}
\usepackage{mathtools}
\usepackage{bbm}
\usepackage{braket}
\usepackage{float}
\usepackage{fancyhdr}
\usepackage{tikz}
\usepackage{calrsfs}
\usepackage{hyperref}
\usepackage{graphicx,epstopdf}
\usepackage{capt-of}
\usetikzlibrary{arrows}
\newcommand{\midarrow}{\tikz \draw[-triangle 90] (0,0) -- +(.1,0);}
\hypersetup{
	colorlinks   = true, 
	urlcolor     = blue, 
	linkcolor    = blue, 
	citecolor   = red 
}

\def\o{\omega}

\def\r{\ref}

\def\a{\alpha}
\def\b{\beta}

\def\g{\gamma}
\def\G{\Gamma}
\def\d{\delta}
\def\D{\Delta}

\def\e{\epsilon}

\def\t{\tau}

\def\f{\frac}

\newcommand{\be}{\begin{equation}}
\newcommand{\ee}{\end{equation}}
\newcommand{\bes}{\begin{equation*}}
\newcommand{\ees}{\end{equation*}}
\newcommand{\rarrow}{\rightarrow}

\def\l{\left}
\def\r{\right}

\NewEnviron{eqn}{
\begin{align}
\begin{split}
  \BODY
\end{split}
\end{align}
}

\NewEnviron{eqn*}{
\begin{align*}
\begin{split}
  \BODY
\end{split}
\end{align*}
}

\newcommand{\ttime}{\tilde{t}}
\newcommand{\tx}{\tilde{x}}

\newcommand{\tvx}{\tilde{\v x}}

\newcommand{\bX}[2]{\bar{X}^{(#1)}\l( #2 \r)}

\newcommand{\intinf}{\int_{-\infty}^{\infty}}

\renewcommand{\v}[1]{\ensuremath{\mathbf{#1}}} 

\newcommand\tab[1][1cm]{\hspace*{#1}}

\begin{document}
\title{ Unruh-DeWitt detector in presence of multiple scalar fields : A Toy Model}
\author{Chandramouli Chowdhury$^{a,b}$\footnote {\color{blue} chandramouli.chowdhury@gmail.com}}
\author{Ashmita Das$^{a}$\footnote {\color{blue} ashmita.phy@gmail.com}}
\author{Bibhas Ranjan Majhi$^{a}$\footnote {\color{blue} bibhas.majhi@iitg.ac.in}}

\affiliation{$^{a}$ Department of Physics, Indian Institute of Technology Guwahati, Guwahati 781039, Assam, India\\
$^{b}$ International Centre for Theoretical Sciences, Bengaluru, North Karnataka 560089, India}



\begin{abstract}
Applications of Unruh-Fulling (UF) effect are well studied in literature via the interaction of Unruh-DeWitt (UD) detector and {\it single} scalar field. In this work, we investigate a toy model, where the detector is interacting {\it simultaneously} with the {\it multiple} scalar fields. Our study reveals that the transition rate of the system significantly depends on the acceleration of the detector and the number of scalar fields ($n$). For $n\gg1$, there exists a {\it critical acceleration}, beyond which the transition rate becomes drastically high than the accelerations below the critical point. The appearance of such critical point never occurs in case of the interaction of UD detector and {\it single} scalar field. 
\end{abstract}
\maketitle
\noindent
\flushbottom
\section{Introduction and Motivation}\label{Introduction}
Since 1973, the theory of Unruh-Fulling (UF) effect \cite{Fulling:1972md,Unruh:1976db} has been of consistent interest to many physicists. 
It describes that an observer traveling with a uniform acceleration $a$ 
through Minkowski vacuum of a quantum field, registers thermal spectrum 
at temperature $T=(\hbar a)/(2\pi c k_B)$ (where $\hbar,~c,~k_B$ is reduced Planck constant,
the speed of light in vacuum, Boltzmann constant respectively), while an inertial observer records no particle excitation in Minkowski vacuum.
This outcome leads us to infer that the concept of particle content in quantum field theory is an observer dependent
phenomenon  \cite{Fulling:1972md,Unruh:1977ga}.  The concept of uniformly accelerated observer is related to uniform gravitational source via the {\it equivalence principle} and therefore after the proposition of UF effect, it was shown that the phenomena of Hawking radiation \cite{Hawking, Unruh:1977ga}, cosmological horizon and particle creation \cite{Gibbons}, particle creation in curved background \cite{Hajicek} can also be perceived by using the UF effect as a tool. 
In this context, it is worth to be mentioned that in \cite{Singleton:2011vh,Crispino:2012zz,Singleton:2012zz}, the authors have studied the response functions of the UD detector in the Schwarzschild spacetime with respect to the Boulware, Unruh, Hartle-Hawking vacua \cite{Birrell} and compared it with the response function of an uniformly accelerated detector in Rindler spacetime, moving with the same locally measured acceleration, with respect to the Minkowski vacuum of the massless scalar field.  It is well known that both the detectors {\it i.e} the detector in curved and as well as flat background, will record thermal radiation from the respective vacuum states. However, it was shown in \cite{Singleton:2011vh,Crispino:2012zz,Singleton:2012zz}, that the temperature as perceived by the detector in the gravitational background is turned out to be higher than the accelerated detector in flat spacetime. This signifies the violation of the equivalence principle. It was also shown that these two temperatures start to become equal if the detector in Schwarzschild spacetime approaches the event horizon {\it i.e} when the gravitational effect gradually becomes stronger and hence the equivalence principle is restored near the event horizon of the black hole.

In terms of the applications of UF effect, the concept of UD detector is indeed a successful technique to realise the thermal spectrum as perceived by the accelerated observer. For some brief review, we refer our readers to \cite{Birrell,Fulling:1987,Takagi,Crispino}. 
In some recent works, the UD detector in the presence of the various type of quantum fields and detector trajectories have been studied, to find its implications in gravitational and cosmological scenarios \cite{Paddy,Oditi,Zhang,Nassif,Sriramkumar,Nicolas}. Also, it has been observed in \cite{Bibhas} that the force experienced by the accelerated observer due to the radiation obeys standard fluctuation-dissipation theorem.

The conventional notion of UD detector dictates that, in flat spacetime, a point like uniformly accelerated detector records particle excitation due to the interaction with a scalar field, located in its Minkowski vacuum. Observing the strong impact of UF effect on the standard results of quantum field theory, a spontaneous question may arise that how a UD detector would behave if it interacts with {\it multiple} quantum fields? This fundamental question is a prime motivation for us to initiate the present work. 

For further motivations, the existence of multiple fields in 4-dimensional spacetime can also be realised in the context of Kaluza-Klein (KK) compactification, where the multiple fields appear as KK modes of a corresponding bulk field \cite{ADD, RS}. 
One can also find in literature \cite{Dvali:2007hz,Dvali:2009ne} that, a large number of particle species have been considered in order to resolve the hierarchy problem of the two fundamental length scales {\it i.e} electroweak and Planck scale. Subsequently, in \cite{Dvali:2009fw}, it was shown that an appropriate abundance of dark matter can be achieved by considering a large number $(N)$ of copies of the standard model along with an imposed additional discrete symmetry, $P(N)$, under which the standard model is charged.

In recent time, the authors of \cite{Alkofer} have studied the modification in Wightman function which appears in the response function of the detector, due to the dimensional flow \cite{Carlip1,Carlip2} and obtained a suppression in the transition rates for the quantum gravity models which exhibit dimensional reduction at high
energies and an increment in the transition rate for a dimensional enhancement at high energies, as in KK models. 
Also in \cite{Chiou}, the response function of a UD detector interacting with the single massless scalar field has been analysed with a compact spatial dimension in D-dimensional flat spacetime. The author of \cite{Chiou} has calculated the response function for the detector moving with uniform acceleration in compact and noncompact dimensions and examined the nonequivalence of inertial, uniform accelerated frames due to the presence of compactified dimension.
Although these studies are highly interesting by themselves, it is on a different footing from our work. 

We clearly state that the purpose of our work is based on the field theoretic interest to understand the behaviour of multi-particle systems, from the perspective of an accelerated observer. 
Therefore to address this issue, we propose a toy model, where the UD detector is interacting {\it simultaneously} with multiple scalar fields, located in their respective Minkowski vacua.
Our work reveals some important aspects as follows:
\begin{itemize}
\item
The transition rate of the system significantly depends on the acceleration of the detector, as well as the number of scalar fields ($n$).
\item
For large $n$ limit, there emerges an important value of the acceleration, which we name as critical acceleration ($a_c$). It is observed that for $a<a_c$, the transition rate is highly suppressed in comparison to the condition $a>a_c$. If one upraises $n$ to very large values, the corresponding transition rates start to exhibit a sharp increment at the critical point $a=a_c$. 
\end{itemize} 
We organise our work as follows.
In sec.(\ref{our work}), we propose a toy model, consists of a point like UD detector and multiple scalar fields and analyse the transition rate of the system. We demonstrate the behaviour of the Wightman function in appropriate limits in sec.(\ref{wightman function}).
The analysis of the transition rate, under the various circumstances, are discussed in sec.(\ref{power spectrum}) and sec.(\ref{F_nvsn}). In sec.(\ref{motivation}), we discuss some of the possible interactions of the UD detector and multiple fields and propose to tackle the nonrenormalisability issue in the context of the present work. Finally, we discuss our results in the last section. Two appendices are being given for the paper to be self-sufficient.
\section{Unruh-DeWitt Detector for Multiple Scalars : The set up of our model}\label{our work}
The interaction of a UD detector and single scalar field is well studied in literature \cite{Birrell,Fulling:1987,Takagi,Crispino} and will not be repeated here.
In the present work, we consider a UD detector interacting simultaneously with the $n$ number of massive real scalar fields, at the same spacetime point in $4$-dimensional Minkowski spacetime. 
The detector moves on a trajectory with uniform acceleration $a$, given by the world line $x(\t)$, where $\t$ is the proper time of the detector. This essentially behaves as a Rindler observer. In the subsequent analysis, we use natural unit for all the parameters, i.e, $\hbar=c=k_b=1$ and metric signature as, $(+, -, -, -)$.
We impose the following conditions in order to keep our analysis simple:
\begin{itemize}
\item
Due to the interaction, all the scalar fields must make a simultaneous transition from their respective ground state to their respective one-particle state while the detector makes a transition from its lower energy eigenstate state $\ket{E_i}$ to a higher energy eigenstate $\ket{E_f}$ [see fig.(\ref{states-2})]. 
\item
The scalar fields are not interacting among themselves and we neglect any entanglement between the vacua and the excited states of the scalar fields. 
\end{itemize}
On the basis of the above conditions and notion of simultaneity, we construct the following structure of interaction Lagrangian for our setup,
\be
\mathcal{L}_{int} =  
\mathfrak{g} \ \mu(\t) \phi_{1}\big[x(\t)\big] \  \phi_{2}\big[x(\t)\big] \  \cdots \ \phi_{n}  \big[x(\t)\big]e^{-s|\t|}~,
\label{lagrangian_multiple}
\ee
The set of initial and final states are, 
  \begin{eqn}
   \ket{\small{\mbox{initial}}} &= \ket{E_i} \otimes \Bigl( \ket{0_1} \otimes \ket{0_2} \otimes   \cdots \otimes \ket{0_n}  \Bigl) \\
   \ket{\small{\mbox{final}}} &= \ket{E_f} \otimes \Bigl( \ket{1_{p_1}} \otimes \ket{1_{p_2}} \otimes \cdots  \otimes \ket{1_{p_n}}\Bigr)~.
  \end{eqn}
  Here, $\ket{0_1}$, $\ket{0_2}$,..,$\ket{0_n}$ represent the Minkowski vacuum states and 
  $\ket{1_{p_1}}$,  $\ket{1_{p_2}}$,.., $\ket{1_{p_n}}$ denote the 1-particle states, for each scalar field. The four momentum associated with each 1-particle state $\ket{1_n}$ is $p_n$.

\subsection*{Transition Amplitude $\rarrow$ Transition Probability}\label{analysis1}
We implement the same procedures of obtaining the response function for single field and UD detector and write the transition amplitude for our case as, 
\begin{eqn}
&A(E_f|E_i)=i\mathfrak{g} \bra{E_f} \mu(0) \ket{E_i} \intinf\,d{\t} e^{i(E_f-E_i)\tau-s|\tau|} \\
&\times \bra{1_{1}} \otimes \cdots \otimes \bra{1_{n}}    \phi_1[x(\t)] \cdots \phi_{n}[x(\t)] \ket{0_{1}} \otimes \cdots \otimes \ket{0_{n}}
\label{amp_multiple1}
\end{eqn}
The transition probability can be obtained by squaring the modulus of amplitude in eq.(\ref{amp_multiple1}) and integrating over all possible momenta for the (relativistically normalised) 1-particle states by using the normalisation, $\int \f{\,d^3 p}{(2\pi)^3} \f{1}{2 E_{p}} \ket{1_p} \bra{1_p} = 1$.
 Therefore, 
\begin{eqn}
&\int \f{\,d^3 p_1}{2E_{p_1}} \cdots \f{\,d^3 p_n}{2E_{p_n}}
|A(E_f|E_i)|^2 =\\
& \mathfrak{g}^2 |q|^2\int_{-\t_0}^{\t_0} \,d\t \,d\t' \ \exp[i \D E (\t - \t')-s(|\t|)-s(|\t'|)]\\
&\times \int \f{d^3 p_1}{2E_{p_1}} \cdots \f{d^3 p_n}{2E_{p_n}} \\
&\times\bra {1_1} \otimes \cdots \otimes \bra{1_n} \phi_1(\t) \cdots \phi_n(\t) \ket{0_1} \cdots \ket{0_n} \\
&\times \bra{0_1} \otimes \bra{0_n} \phi_1(\t') \cdots \phi_n(\t') \ket{1_1} \otimes \cdots \ket{1_n}\label{trans_prob}
\end{eqn}
Here, $\D E = (E_f - E_i)$. Dividing the above equation by $2\t_0$, which is the duration of interaction, the transition rate becomes,
\begin{eqn}
&P(E_f|E_i) \\
&= \f{\mathfrak{g}^2|q|^2}{2\t_0} \int \f{\,d^3 p_1}{2E_{p_1}} \cdots \f{\,d^3 p_n}{2E_{p_n}} |A(E_f|E_i)|^2 \\
&= \f{\mathfrak{g}^2|q|^2}{2\t_0} \int_{-\t_0}^{\t_0}\,d{\t} \,d{\t'} e^{i \D E (\t - \t')} \\ 
&\times\bigg[ \bra{0_1} \phi_1(\t) \phi_1(\t') \ket{0_1} \cdots \bra{0_n} \phi_n(\t) \phi_n(\t') \ket{0_n} \bigg]\label{transition rate_multiple}
\end{eqn}
We stretch the duration of interaction to infinity (i.e,  $\t_0\rightarrow \infty$) and the adiabatic factor $\mathfrak{s}\rightarrow 0$ in the above expression and obtain,
\be
R(E_f|E_i) =\lim_{\substack
{s\to 0 \\
\tau_0\to\infty}}P(E_f|E_i)= |q|^2 F_n(\D E),
\label{rate_multiple1}
\ee
with $q = \bra{E_f} \mu(0) \ket{E_i}$ and 
\begin{equation}
F_n(\D E)=\mathfrak{g}^2\intinf \,d{\t}\  e^{-i \D E \t} \  \prod_{i = 1}^n \D_i(\t)
\label{response_multiple1}
\end{equation}
Here $F_n(\D E)$ represents the response function of the detector while interacting with $n$ number of scalar fields and the Wightman functions are denoted by,
\begin{eqn}
\D_i(\t) = \bra{0_i}{\phi_i(x(\bar\t + \t)] \ \phi_i[x(\bar\t)]}\ket{0_i}\label{wightman}
\end{eqn}
\section{Unwrapping the Wightman Function}\label{wightman function}
We concentrate
on the form of Wightman function in the current section. The standard expression
for the Wightman function of a massive scalar field in
$4$-dimensional Minkowski spacetime can be written as \cite{Birrell,Alkofer},
\be
\D_i(x - x') = -\f{i m_i}{4\pi^2} \cdot \f{K_1\bigg( i m_i \sqrt{(t - t' - i \e)^2 - (\v x - \v x')^2} \bigg)}{\sqrt{(t - t')^2 - (\v x - \v x')^2}}
\label{wight_multiple1}
\ee
Here, $K_{\nu}(z)$ is the modified Bessel function. The relation between the modified Bessel function and the Bessel function of first kind is explicitly given in appendix \ref{Analysis of Bessel Function}. The parameters $(t,\v{x})$ in eq.(\ref{wight_multiple1}) are the usual Minkowski spacetime coordinates. In order to get the form of Wightman function from the perspective of a Rindler observer, one needs to transform the Minkowski coordinates in eq.(\ref{wight_multiple1}) to Rindler coordinates by using Rindler transformation.

The Rindler coordinates $(\eta, \xi, x_{\perp})$ are related to Minkowski coordinates by,
\begin{equation}
t=\xi \sinh{\eta}, \tab x=\xi  \cosh{\eta}
\label{rindler1}
\end{equation}
We denote the transverse coordinates $(y,z)$ by $x_{\perp}$, which are unchanged under the Rindler transformation.
Considering, $\xi(\tau)=a^{-1}$ and $\eta(\tau)=a\tau$, the world line of the detector becomes,
\begin{eqn}
&t(\t) = a^{-1} \sinh(a \t), \tab x(\t) = a^{-1} \cosh(a\t), \\
&x_{\perp}(\t) = \mbox{constant}
\label{rindler2}
\end{eqn}
We denote, $s = \sqrt{\ttime^2 - \tvx^2}$, where any parameter $\tilde{Q} = Q - Q'$.
Using eq.(\ref{rindler2}), the parameter $s$ becomes,
\be
s^2=\f{4}{a^2}\sinh^2\l(\f{a \t}{2}\r)
\label{rindler trajectory}
\ee
The above choice of the trajectory of the detector signifies that the detector is essentially moving in a hyperbolic trajectory in the Right Rindler Wedge, which implies the detection of quanta of positive frequencies only.

In order to compute the response function we consider the ``{\textit{small}}'' and ``{\textit{large argument}}'' limit of the Wightman function in eq.(\ref{wight_multiple1}). Upon using eq.(\ref{rindler trajectory}) the argument of the Bessel Function becomes [$(2im/a) \sinh(a\t/2)$]. The large and small argument limit implies the expansion with respect to the value of the dimensionless parameter $m/a$. The expansions of the modified Bessel's function with respect to the small/large argument is elaborately given in appendix \ref{Analysis of Bessel Function}.
The small argument expansion can be written as \cite{Takagi}, (we assume that all the scalar fields are of the same order of masses and therefore now onwards we suppress the index ``$i$'')
\begin{eqn}
\D(\tx) = -\f{1}{4 \pi^2 s^2} + \f{m^2}{8\pi^2} {\rm ln}\bigg(\frac{m}{a}\bigg) + \f{m^2}{16 \pi^2} \big[ \d +  {\rm ln}(isa) \big] 
\label{low mass}
\end{eqn}
Here, $\d$ is a constant with the value $(2 \g - 1 - \log{4})$ with $\g$ being the Euler-Gamma constant. Similarly a large argument expansion of eq.(\ref{wight_multiple1}) can be written as \cite{Takagi}, 
\be
\D(\tx) = \f{1}{4 \pi s} \ \l( \f{im}{2 \pi s} \r)^{1/2} e^{-i m s}
\label{high mass}
\ee
We mention that, one can in principle, expand the field in terms of the Rindler modes also. These two procedures of quantisation (Minkowski/Rindler quantisation) produces the same results for any physical observable, such as the power spectrum (or) the transition rate \cite{Crispino,Takagi}. 
\section{Evaluation of Transition Rate}\label{power spectrum}
In our case, a product of Wightman functions (as in eq.(\ref{low mass} and \ref{high mass})) appear for $n$ number of scalar fields in the expression for transition rate ({\it i.e} eq.(\ref{response_multiple1})).
For, $m/a \ll 1$,  one can clearly understand that the product of Wightman functions is very complicated to evaluate, due to the presence of the third term in eq.(\ref{low mass}). A proper series expansion in $m/a$ is not feasible because of the interference of the second and third term in eq.(\ref{low mass}), which are of the same order. Therefore we perform our rest of the analysis of the low argument limit of the Bessel function by considering the massless limit of the scalar fields for the sake of simplicity.

On the other hand, the large argument limit of the Bessel function as in eq.(\ref{high mass}) produces a more tractable form in the final expression for the transition rate, in comparison to the low argument limit.
\subsection{Case I: $\v{m\rightarrow 0}$}\label{small mass}
For $n$ number of massless scalar fields, the product of Wightman functions {\it i.e},  $\D_1(x, x') \D_2(x, x') \cdots \D_n(x, x')$, becomes, 
\begin{eqn}
\l(-\f{1}{4\pi^2}\r)^n &\bigg[  \f{1}{s^{2n}} \bigg]
\end{eqn}
We denote, $\D E=\o$ and therefore the response function becomes,
\be
F_n(\o) = \l( - \f{1}{4 \pi^2}\r)^n \mathfrak{g}^2 \intinf\,d{\t} \ e^{-i \o t} \ \bigg[ \f{1}{s^{2n}} \bigg]\label{multiple_power2}
\ee
 We write eq.(\ref{multiple_power2}) as,
\begin{eqn}
F_n(\o) = &\l(- \f{1}{4\pi^2} \r)^n \mathfrak{g}^2\ \l(\f{a}{2}\r)^{2n}
\times I(\o, a, n) ~
\label{large n power}
\end{eqn}
where, 
\be
I(\o, a, n) = \intinf\,d{\t} \ \f{e^{-i \o \t}}{\sinh^{2n}\l( a \t/2 \r)}~
\label{large n power1}
\ee
The above integral turns out to be \cite{Gradshteyn}, 
\begin{eqn}
&I(\o, a, n)\\
&= \l(\f{2\pi}{\o}\r)\f{(-1)^n}{(2n-1)!} \f{4^n}{e^{2\pi\o/a} -1}\prod_{k=1}^{n} \bigg[ \f{\o^2}{a^2} + (n-k)^2\bigg]~
\label{Integral I}
\end{eqn}
Using eq.(\ref{Integral I}), the response function of the detector becomes, 
\begin{eqn}
F_n(\o) &= (-1)^{2n} \mathfrak{g}^2\l( \f{2\pi}{\o}\r) \f{1}{e^{2\pi\o/a} -1}\l( \f{a}{2\pi}\r)^{2n} \\
&\times\bigg[  \f{1}{(2n-1)!} \prod_{k = 1}^n \l\{ \f{\o^2}{a^2} + (n - k)^2 \r\} \bigg]
\label{multiple_power3}
\end{eqn}
The above equation can also be expressed as,
\begin{eqn}
F_{n}(\o) = n \mathfrak{g}^2&\l( \f{a}{2\pi} \r)^{2n} \l( \f{2\pi}{\o} \r) \f{1}{e^{2\pi\o/a} - 1}\\
&\times\bigg[ \f{(n-1)!^2}{(2n-1)!} \bX{n-1}{\f{\o}{a}} \bigg]~
\label{multiple_power4}
\end{eqn}
where the function $\bX{n}{\o/a}$ is a modification of the Pocchammer function whose explicit expression is given by eq.(\ref{pocchammer}) of appendix \ref{large_n}.

Now we make a comment on the dimension of the coupling of the interaction. We concentrate only to the dimension-full terms of eq.(\ref{multiple_power3}), and therefore, the transition rate can be written as, 
\be
R(E_f|E_i) \sim \mathfrak{g}^2 q^2 \ \f{a^{2n}}{\o}~
\ee
In literature, we find that one usually defines the transition rate by rescaling the quantity $R(E_f|E_i)$ by $q^2$. In our case we stick to the same convention which implies that the transition rate becomes equal to $F_n(\o)$ [see eq.(\ref{rate_multiple1})]. Therefore we use the terminology of transition rate of a system and the response function/power spectrum of the detector interchangeably. 
\be
\f{R(E_f|E_i)}{q^2} =F_n(\o) \sim \mathfrak{g}^2 \ \f{a^{2n}}{\o}~
\label{power dim}
\ee
It is well known that in standard UD detector and single field case, the response function turns out to be : $F_{standard}(\o) = \f{\o}{e^{2\pi\o/a}-1}$ and possesses the mass dimension $[M]^1$ \cite{Takagi,Crispino}. In this expression for $F_{standard}(\o)$ all the parameters carry the same notion (as described in sec(\ref{our work})) for the UD detector and single field scenario. Therefore, from eq.(\ref{power dim}), the mass dimension of $\mathfrak{g}$ turns out to be, $[M]^{1-n}$.
Hence, we re-parametrise $\mathfrak{g}$ in terms of $\a$ as,
\be
\mathfrak{g} = \a^{1 - n}~
\label{g M rel}
\ee 
Where $\a$ is a parameter with mass dimension $[M]^{1}$. 
We choose such parametrisation in order to write the expression for transition rate in a more compact form and maintain the correct dimensionality of $F_n(\o)$. Hence eq.(\ref{multiple_power4}) can be written as,
\begin{eqn}
F_n(\o)=n\a &\bigg(\f{2\a \pi}{\o}\bigg)\f{1}{e^{2\pi\o/a}-1} \l(\f{a}{2\pi\a}\r)^{2n}\\
&\times \bigg[  \f{(n-1)!^2}{(2n-1)!} \bX{n-1}{\f{\o}{a}} \bigg]
\label{trans_multiple1}
\end{eqn}
\subsection{Case II: $\v{m/a \gg 1}$}\label{large mass}
In the large argument limit, using eq.(\ref{high mass}), the response function of the detector, interacting with $n$ number of scalar fields can be written as, 
{\small\be
F_n(\o) = \mathfrak{g}^2\l(\f{1}{4\pi^2} \r)^n  \l(\f{i}{2 \pi}\r)^{n/2} \intinf \,d{\t} e^{-i \o \t} \f{\exp\big[-is \Sigma_m\big]}{s^{3n/2}}
\label{response_masive}
\ee}
Here, $\Sigma_m \equiv \sum_{i=1}^n m_i$ and $m_i$ denotes the mass of the $i^{th}$ scalar field. We examine
the power spectrum as in eq.(\ref{response_masive}), by revisiting the proof as given in $\S 4$ of \cite{Takagi}.
Therefore one can see that the Wightman function for $m/a\gg1$ varies as [see eq.($4.5.3$) in \cite{Takagi}], 
\be
|g_n(\t)| \sim \f{1}{|s|^{3n/2}} \exp[- \e \ \Sigma_m]
\label{wight_multiple}
\ee
Here we add a small positive imaginary part, $\e$ to the interval $s$, to handle the divergence at $\t = 0$. Thus, 
\be
s =  \f{2}{a} \sinh\l(\f{a}{2}\t\r) + i \e
\label{regulated s}
\ee
Therefore the transition rate can be written as, 
\be
F_n(\o) = \mathfrak{g}^2\int_{-\infty}^{\infty} \,d{\t} \ e^{-i \o \t} g_n(\t)
\ee
At this point, we emphasise that the above integral is over the variable $\t$, where $-\infty<\t<\infty$, whereas in this case, the quantity $\f{m}{a}$ is always considered to be $\gg 1$. We are specially interested in the large argument limit of the Wightman function, hence, 
\be
\lim_{m \to \infty} F_n(\o) \equiv \mathfrak{g}^2\lim_{m \to \infty} \int_{-\infty}^{\infty} \,d{\t} \ e^{-i \o \t} g_n(\t)
\label{convergence_largemass1}
\ee
By using the Dominated Convergence Theorem (Lebesgue's theorem) to the above equation, it can be argued that this integral vanishes. We note that from eq.(\ref{wight_multiple}), it is easy to convince oneself that, for large enough $\Sigma_m$ (where the bound is strengthened as $m \to \infty$) one can always find a function $\bar{g_n}(\t)$ independent of $\Sigma_m$, which satisfies the conditions, $\bar{g_n}(\t) > |g_n(\t)|$ and $\intinf \,d{\t} \ \bar{g_n}(\t) < \infty$.
These allow one to use the Lebesgue's convergence theorem [see \cite{Halmos} $\S 26$] to push the limit inside the integral in the RHS of eq.(\ref{convergence_largemass1}) and conclude, 
\be
\lim_{m \to \infty} F_n(\o) = \mathfrak{g}^2 \int_{-\infty}^{\infty} \,d{\t} \ e^{-i \o \t} \lim_{m \to \infty} g_n(\t) = 0
\ee
Therefore, we obtain that in the large mass limit of the scalar fields the power spectrum of the detector approaches to zero. This result is same as the standard case of single massive scalar field and detector interaction which is elaborately discussed in \cite{Takagi} (section $\S4.5$).
\section{Analysis of Transition Rate}\label{F_nvsn}
We devote this section to study the behaviour of power spectrum of the detector with respect to the various parameters of our theory. Since the power spectrum is almost negligible for the case $m/a\gg 1$, the region of interest lies in $m\rightarrow 0$ limit. Thus we restrict ourselves to massless parametric regime to study the behaviour of transition rate. 

Eq.(\ref{trans_multiple1}) reveals that the response function of the UD detector depends on the parameters -- $n,~\o,~a,~\a$. To begin with, we plot the power spectrum {\it i.e} $F_n(\o)$ as a function of $\o$ for fixed values of $n, a$ and $\alpha$. In this analysis $a,~\o,~\a$ are in eV mass unit. 
\begin{figure}[H]
\includegraphics[width=3.0in,height=1.74in]{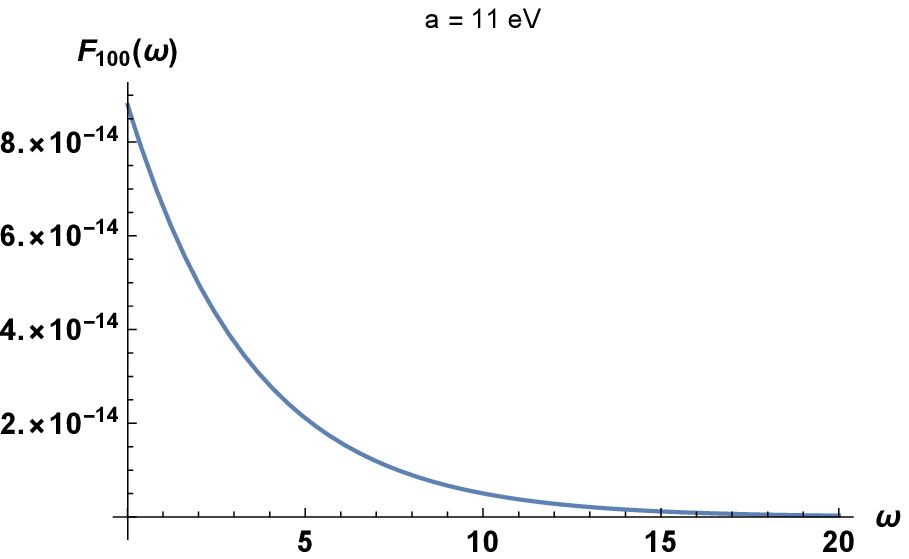}
\caption{$F_n(\o)$ vs $\o$ for $n=100$, $\a=1$ eV}
\label{a<a_cplot}
\end{figure}
\begin{figure}[H]
\includegraphics[width=3.0in,height=1.74in]{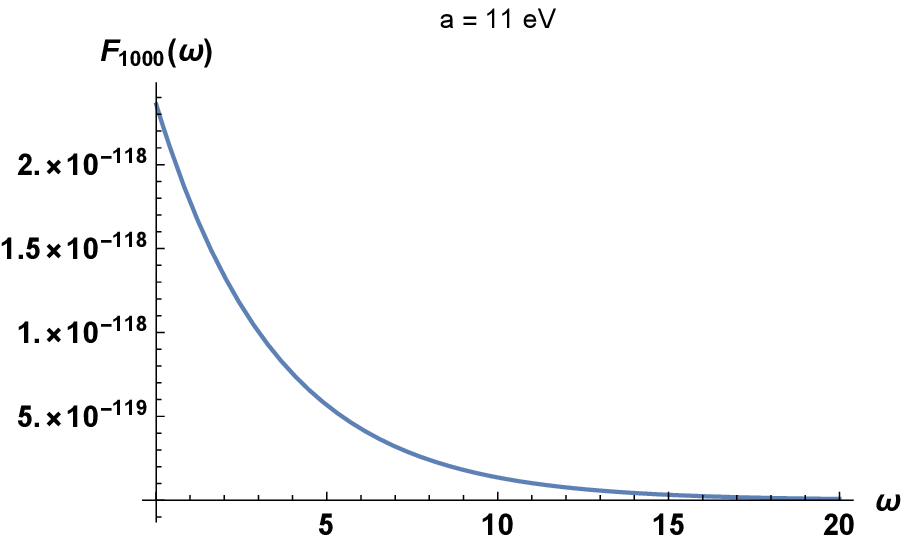}
\caption{$F_n(\o)$ vs $\o$ for $n=1000$, $\a=1$ eV}
\label{a<a_c_1000plot}
\end{figure}
\begin{figure}[H]
\includegraphics[width=3.0in,height=1.74in]{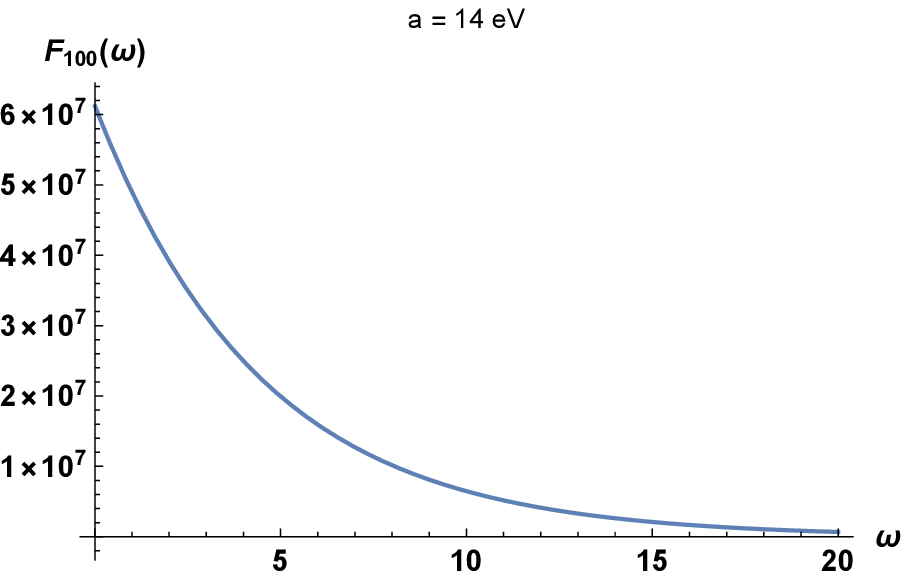}
\caption{$F_n(\o)$ vs $\o$ for $n=100$, $\a = 1$ eV}
\label{a>a_cplot}
\end{figure}
\begin{figure}[H]
\includegraphics[width=3.0in,height=1.74in]{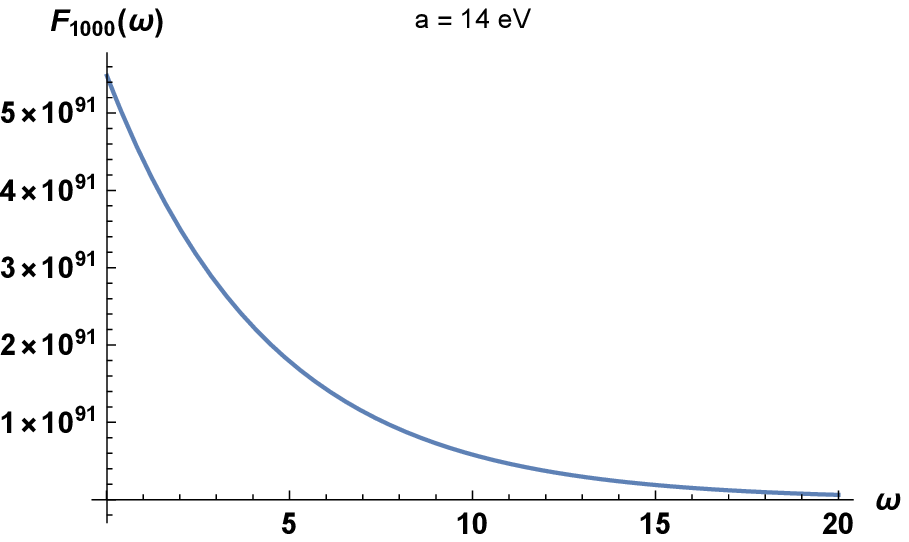}
\caption{$F_n(\o)$ vs $\o$ for $n=1000$, $\a = 1$ eV}
\label{a>a_c_1000plot}
\end{figure}
We describe the results, obtained by the above graphical analysis, below:
\begin{itemize}
\item It is evident from the above plots that the transition rate decreases with the increasing value of $\o$ and gradually approaches to zero. This converging nature of the power spectrum with increasing $\o$ is similar to the case of the single massless scalar field and UD detector.
One can understand this feature of convergence, by examining the function,
\bes
\f{\bX{n-1}{\o/a}}{e^{2\pi\o/a} - 1}
\label{convergence_omega}
\ees
The explicit form of the function $\bX{n}{\a}$ is given in eq.(\ref{pocchammer}). This function has a bell-like nature (or) an exponentially decaying nature when varied with respect to $n$ for all finite values of $\o$. Whereas this function is not very sensitive to the value of $a$. One can see that for large $\o$ the exponential factor in the denominator dominates the numerator and hence justifies the convergence of $F_n(\o)$. 
\item
Fig.(\ref{a<a_cplot}) and fig.(\ref{a<a_c_1000plot}), depict that the transition rates turn out to be very small for different $n$ values. Using the numerical values of all the parameters as specified in these two figures and fixing $\o=5$ eV, one obtains $F_n(\o)\sim O(10^{-13})$ eV and $O(10^{-119})$ eV respectively from eq.(\ref{trans_multiple1}).
\item
In fig.(\ref{a>a_cplot}), fig.(\ref{a>a_c_1000plot}), for $a=14$ eV, one finds that the transition rate of the detector is drastically enhanced for both the values of $n$, in contrast to the transition rate for $a=11$ eV.
The value of $F_n(\o)$ corresponding to $\o=5$ eV, turns out to be $O(10^{8})$ eV and $O(10^{91})$ eV respectively.
\end{itemize}
This analysis indicates that there exists a definite value of acceleration beyond which the transition rate of the detector starts to increase significantly. We name this as {\it critical acceleration} $a_c$. We also notice that for $n=1000$ the change in the transition rate is more drastic than $n=100$ when the acceleration $a$ crosses the value $a_c$.

Now we study the transition rate in large $n$ limit by using eq.(\ref{trans_multiple1}). We use Stirling's approximation in eq.(\ref{trans_multiple1}) to the ratio of factorials (see Appendix \ref{large_n}, eq.(\ref{stirlingapprox1})). Moreover it is evident from eq.(\ref{pocchammer}), that under the condition $n\gg1$, the term $\bX{n-1}{\f{\o}{a}}$ produces some constant number. Therefore for large $n$ limit the power spectrum in eq.(\ref{trans_multiple1}) takes the form:
\begin{equation}
F_{n \gg 1}(\o)\cong \a^2\ n^{-1/2}\bigg(\f{a}{4\pi\a}\bigg)^{2n}\f{2\pi^{3/2}}{\o}\f{1}{e^{2\pi\o/a}-1}
\label{n_infinity_trans1}
\end{equation}

From the above equation, It can be realised analytically that for $n \gg 1$ and for finite $\o$, the factor $\l(\f{a}{4\pi \a}\r)^{2n}$ governs the order of magnitude of $F_n(\o)$. Therefore, it can be argued, that the transition rate exhibits a drastic enhancement with the increasing number of fields for $a> 4\pi\a$, which allow us to identify the factor $4\pi\a$ as the critical acceleration $a_c$. Considering larger and larger values of $n$, one can realise that the transition rate undergoes a steep increment, with the variation of acceleration from $a<a_c$ to $a>a_c$.

In order to visualise the variation of the transition rate with respect to both the parameters $\o$, $a$, we produce a single contour plot by fixing $n$, $\a$ in the following figure.
\begin{figure}[H]
\includegraphics[width=3.1in,height=2.4in]{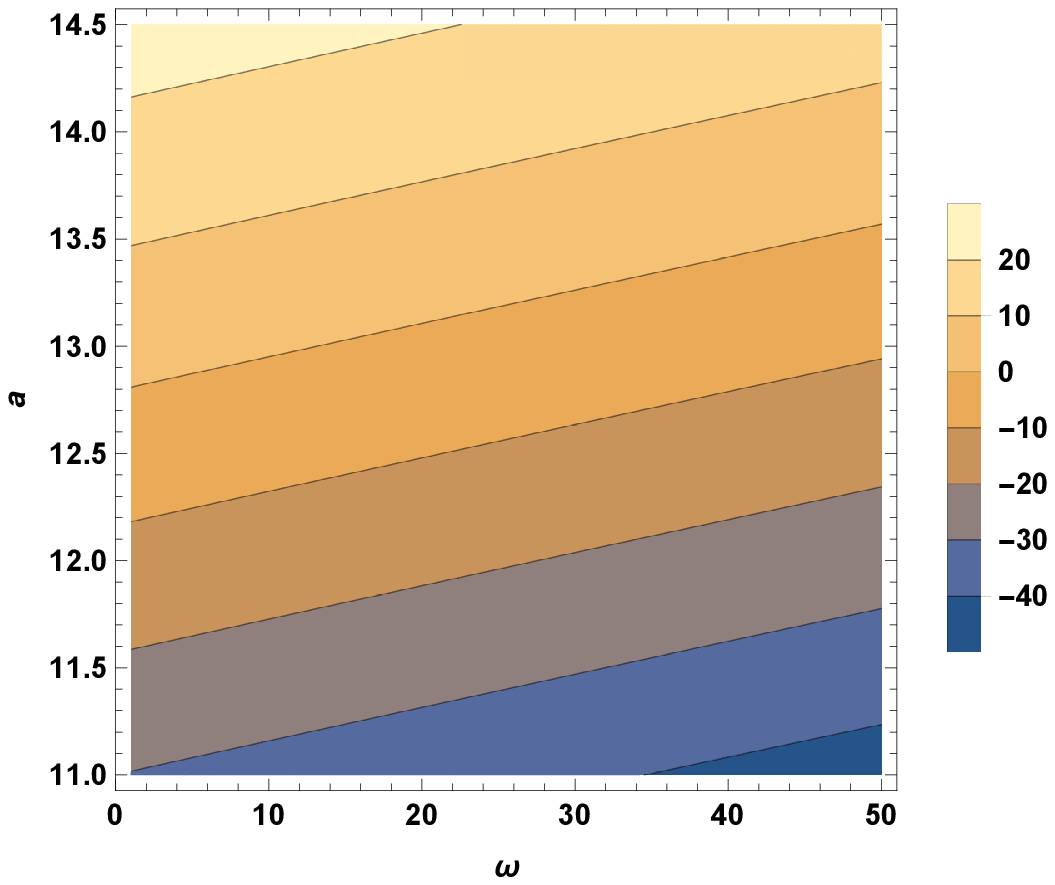}
\caption{$\log[F_n(\o)]$ vs $(\o, a)$ as $a$ goes from $a_c^-$ to $a_c^+$. 
with $n=100$, $\a=1$ eV}
\label{a3dplot}
\end{figure}
It can be seen that these regions are increasing linearly (indicating an exponential increase) while we trace the surface along the increasing acceleration $a$. On the other hand, for a fixed value of acceleration a particular surface is decreasing with increasing value of $\o$, which justifies the convergence of the power spectrum with increasing $\o$ as depicted in fig.(\ref{a<a_cplot})-fig.(\ref{a>a_c_1000plot}). This plot, also brings out the value of critical acceleration, which can be approximately identified to be the point where the log of the transition rate hits zero.

In this context, we mention that the enhancement of the response function of the UD detector along with the increment of the temperature of the thermal bath, have been reported earlier in literature \cite{Bell:1982qr,Akhmedov:2006nd}. In these references, the authors have examined the possibility of obtaining the experimental signature of the UF effect by studying the response function of the detector. However, these studies are concerning with the interaction of single quantum field and UD detector, executing a nonlinear {\it i.e} circular motion, instead of a linear motion as in the conventional one.

\begin{figure}[H]
\includegraphics[scale=0.8]{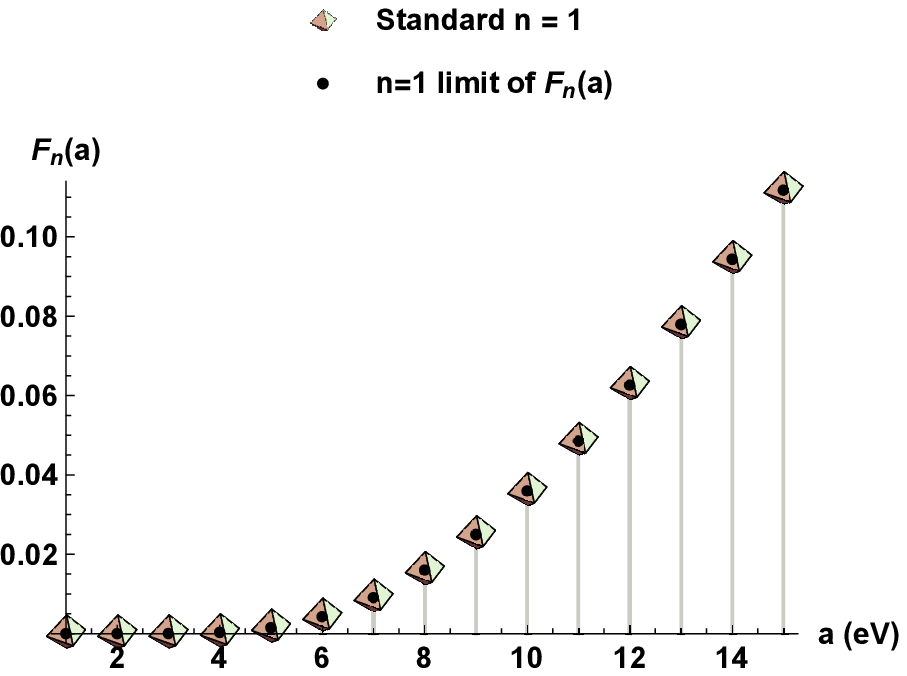}
\caption{$F_n(a)$ vs $a$, for $\o=5$ eV, $n=1$, $\a=1$ eV}
\label{single_check_plot}
\end{figure}

At this stage, we urge to compare the single field limit of our results with the standard UD detector and single field case \cite{Takagi,Crispino}. Therefore we plot $F_{standard}(\o)$ (as specified in sec.(\ref{power spectrum})) and eq.(\ref{trans_multiple1}) jointly in fig.(\ref{single_check_plot}), as a function of $a$ while fix the parameters $\o,\a$.
The above figure depicts that the $n=1$ limit of generalised transition rate ({\it i.e},  eq.(\ref{trans_multiple1})), exactly reproduces the transition rate of the standard UD detector and single massless scalar field, as obtained in literature.
\section{Discussions on the interaction Lagrangian}\label{motivation}
The standard interaction term of a UD detector and single scalar field is of the form, $\sim \mu({\t}) \phi[x(\t)]$, which represents a monopole like interaction. In this case, the detector makes a transition from its lower energy eigenstate $\ket{E}_i$ to the higher energy eigenstate $\ket{E}_f$(representing a ``detection'' of a ``quanta'' of the field), whereas the field makes a transition from its ground state to its 1-particle state (i.e, $\ket{1_p}$, representing the ``quanta'').
\begin{figure}[h]
 \begin{center}
  \begin{tikzpicture}
   
   \begin{scope}[thick, every node/.style={sloped,allow upside down}]
      \draw[purple] (-0.5,0) -- (0.5,0);
      \draw[teal] (-0.5,2) -- (0.5,2);
      \node at (0,-0.25) {\large{$\phi_i$}};
      \draw (0,0) -- node{\midarrow} (0,2);
      \node at (-0.75, 0) {$\ket{0}$};
      \node at (-0.75, 2) {$\ket{1}$};
   \end{scope}

   \begin{scope}[shift = {(-2, 0)}, thick, every node/.style={sloped,allow upside down}]
      \draw[purple] (-0.5,0) -- (0.5,0);
      \draw[teal] (-0.5,2) -- (0.5,2);
      \node at (0,-0.25) {\large{$\mu$}};
      \draw[brown] (0,0) -- node{\midarrow} (0,2);
      \node at (-0.75, 0) {$E_i$};
      \node at (-0.75, 2) {$E_f$};
   \end{scope}
      
  \end{tikzpicture}
 \end{center}
           \caption{Transition of UD detector and single scalar field from their respective initial state to final state}
            \label{standard_unruh}
\end{figure}
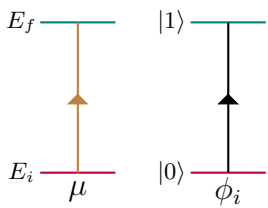
  The transition of the detector and the scalar field from their respective initial state to final state can diagrammatically be represented as in the fig.(\ref{standard_unruh}). Allowing upto the first order in perturbation theory, the transition amplitude of the above system can be written as, 
  \be
  A(E_f|E_i) \sim \bra{E_f} \mu(\t) \ket{E_i} \bra{1_p} \phi[x(\t)] \ket{0}
  \label{standard amplitude}
  \ee

 It is worth to be mentioned at this point that, in the standard notion of UD detector and the single field interaction, the transition amplitude is written up to the first order in perturbation theory, which leads to the fact that, the scalar field must make a transition from its vacuum to 1-particle state, in order to obtain a non-zero transition amplitude. However if one urges to consider the field transition to be a vacuum to 2-particle state, the non-zero transition amplitude would appear in the second order in perturbation theory and so on.

As in this present article, we are addressing a spontaneous question that ``How would a UD detector interact, if there are multiple fields present in the system?'' For multiple scalar fields, we encounter several choices, both, at the level of the choice of final states and the structure of the interaction Lagrangian. This freedom of choices of various final states and interaction Lagrangian are some of the crucial distinctions between the single field and multiple fields interaction with UD detector. The thorough analysis of these possible interactions would be extremely useful and may have their own merits in order to understand various possible aspects related to the studies of observer dependent quantum effects.

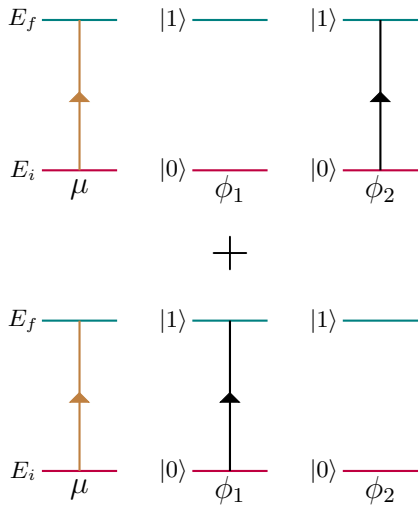
\begin{figure}[h]
 \begin{center}
  \begin{tikzpicture}[scale = 1]
   
   \begin{scope}[thick, every node/.style={sloped,allow upside down}]
      \draw[purple] (-0.5,0) -- (0.5,0);
      \draw[teal] (-0.5,2) -- (0.5,2);
      \node at (0,-0.25) {\large{$\phi_2$}};
      \draw (0,0) -- node{\midarrow} (0,2);
      \node at (-0.75, 0) {$\ket{0}$};
      \node at (-0.75, 2) {$\ket{1}$};
   \end{scope}
   
   \begin{scope}[shift = {(-2, 0)}, thick, every node/.style={sloped,allow upside down}]
      \draw[purple] (-0.5,0) -- (0.5,0);
      \draw[teal] (-0.5,2) -- (0.5,2);
      \node at (0,-0.25) {\large{$\phi_1$}};
      \node at (-0.75, 0) {$\ket{0}$};
      \node at (-0.75, 2) {$\ket{1}$};
   \end{scope}

   \begin{scope}[shift = {(-4, 0)}, thick, every node/.style={sloped,allow upside down}]
      \draw[purple] (-0.5,0) -- (0.5,0);
      \draw[teal] (-0.5,2) -- (0.5,2);
      \node at (0,-0.25) {\large{$\mu$}};
      \draw[brown] (0,0) -- node{\midarrow} (0,2);
      \node at (-0.75, 0) {$E_i$};
      \node at (-0.75, 2) {$E_f$};
   \end{scope}
   
   \node at (-2, -1.1) {\huge{$+$}};;
   
   \begin{scope}[shift = {(0, -4)}, thick, every node/.style={sloped,allow upside down}]
      \draw[purple] (-0.5,0) -- (0.5,0);
      \draw[teal] (-0.5,2) -- (0.5,2);
      \node at (0,-0.25) {\large{$\phi_2$}};
      \node at (-0.75, 0) {$\ket{0}$};
      \node at (-0.75, 2) {$\ket{1}$};
   \end{scope}
   
   \begin{scope}[shift = {(-2, -4)}, thick, every node/.style={sloped,allow upside down}]
      \draw[purple] (-0.5,0) -- (0.5,0);
      \draw[teal] (-0.5,2) -- (0.5,2);
      \node at (0,-0.25) {\large{$\phi_1$}};
      \draw (0,0) -- node{\midarrow} (0,2);
      \node at (-0.75, 0) {$\ket{0}$};
      \node at (-0.75, 2) {$\ket{1}$};
   \end{scope}

   \begin{scope}[shift = {(-4, -4)}, thick, every node/.style={sloped,allow upside down}]
      \draw[purple] (-0.5,0) -- (0.5,0);
      \draw[teal] (-0.5,2) -- (0.5,2);
      \node at (0,-0.25) {\large{$\mu$}};
      \draw[brown] (0,0) -- node{\midarrow} (0,2);
      \node at (-0.75, 0) {$E_i$};
      \node at (-0.75, 2) {$E_f$};
   \end{scope}

  \end{tikzpicture}
 \end{center}
           \caption{Transition of UD detector and two scalar fields from their respective initial state to final state}
           \label{2field-1}
 \end{figure}

Now, for illustration, we consider the interaction of a UD detector and two scalar fields, where we discuss some of the possible choices of the interaction Lagrangian and the final states of the system.
  The most intuitive interaction Lagrangian that one can consider, be written as, 
  \be
 \mathfrak{L}_{int} =\mathbbm{g}\, \mu(\t)(\phi_1[x(\t)]+ \phi_2[x(t)])~,
 \label{scalar-2}
  \ee  
where $\mathbbm{g}$ and $\mu({\t})$ are the coupling strength of the detector with the scalar fields and monopole moment, respectively.  In the presence of multiple fields, one of the possible transition of the system will resemble with the fig(\ref{2field-1}), 
where the the initial and the final states of the detector and the scalar fields can be written as following,
\begin{eqn}
\ket{\small{\mbox{initial}}} &= \ket{E_i} \otimes \ket{0_1} \otimes \ket{0_2} ~;\\
\ket{\small{\mbox{final}}} &= \ket{E_f} \otimes \Bigl( \ket{1_{p_1}} \otimes \ket{0_2}  +  \ket{0_1} \otimes \ket{1_{p_2}}  \Bigr)~.
\label{states-1}
\end{eqn}
The subscripts on the vacuum and one-particle states are indicating the corresponding vacuum and 1-particle state
for each scalar field where  $p_1, \, p_2$ are the four momentum associated with
each 1-particle state of the scalar field. Stating it in a slightly more verbose manner: in this scenario, we have a detector interacting with a set of scalar fields where only a single field interacts with the detector at a particular instant. This set of initial and final state will result in a transition amplitude which simply varies as $\sim (A_1 + A_2)$, with $A_1,\, A_2$ denoting the amplitudes in eq.(\ref{standard amplitude}), due to the interaction of the corresponding scalar fields $\phi_1, \,\phi_2$ with the UD detector. Such of a final state seems to be simple at the level of the evaluation of transition amplitudes, but leads to certain complications when one proceeds onto evaluating the response function which varies as $\sim \sum_{\tiny\mbox{final}} |A_1 + A_2|^2$. Here the summation heuristically denotes a summation on the final set of states of the system, which is in this case, the momenta of the 1-particle states of the scalar fields. This quantity inevitably leads to ``cross-terms'' which look like, 
\be
\int dp_1 \ dp_2 \Bigl[ A_1(p_1, \t) A^*_2(p_2, \t') + A^*_1(p_1, \t) A_2(p_2, \t')  \Bigr]~.
\ee
Contributions from such terms have to be examined more deeply, and currently, we have reserved this issue as a potential future work.

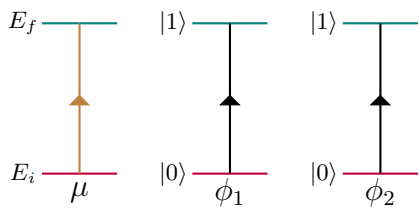
\begin{figure}[h]
 \begin{center}
  \begin{tikzpicture}[scale = 1]
   
   \begin{scope}[thick, every node/.style={sloped,allow upside down}]
      \draw[purple] (-0.5,0) -- (0.5,0);
      \draw[teal] (-0.5,2) -- (0.5,2);
      \node at (0,-0.25) {\large{$\phi_2$}};
      \draw (0,0) -- node{\midarrow} (0,2);
      \node at (-0.75, 0) {$\ket{0}$};
      \node at (-0.75, 2) {$\ket{1}$};
   \end{scope}
   
   \begin{scope}[shift = {(-2, 0)}, thick, every node/.style={sloped,allow upside down}]
      \draw[purple] (-0.5,0) -- (0.5,0);
      \draw[teal] (-0.5,2) -- (0.5,2);
      \node at (0,-0.25) {\large{$\phi_1$}};
      \draw (0,0) -- node{\midarrow} (0,2);
      \node at (-0.75, 0) {$\ket{0}$};
      \node at (-0.75, 2) {$\ket{1}$};
   \end{scope}

   \begin{scope}[shift = {(-4, 0)}, thick, every node/.style={sloped,allow upside down}]
      \draw[purple] (-0.5,0) -- (0.5,0);
      \draw[teal] (-0.5,2) -- (0.5,2);
      \node at (0,-0.25) {\large{$\mu$}};
      \draw[brown] (0,0) -- node{\midarrow} (0,2);
      \node at (-0.75, 0) {$E_i$};
      \node at (-0.75, 2) {$E_f$};

   \end{scope}

  \end{tikzpicture}
 \end{center}
            \caption{Simultaneous transition of UD detector and two scalar fields from their respective initial state to final state}
            \label{states-2}
 \end{figure}

 We now move onto the other possible final state configuration, where both the fields make a simultaneous transition due to the interaction with the UD detector at the same point in spacetime. We refer our readers to the fig.(\ref{states-2}) for a more transparent understanding.

 From the above diagram, it can be seen that the final state of the system becomes a tensor product between all the one-particle states of the scalar fields,
 \be
 \ket{\small{\mbox{final}}} = \ket{E_f} \otimes \ket{1_{p_1}} \otimes \ket{1_{p_2}} ~.
\label{final_2}
 \ee
It is immediately clear that in order to observe a non-zero transition amplitude, with respect to the interaction Lagrangian as in eq.(\ref{scalar-2}) and the initial, final states as written in eq.(\ref{states-1}) and eq.(\ref{final_2}) respectively, we need to go to a higher order term in perturbation theory, i.e, $\sim \bra{\small{\mbox{final}}} \mathfrak{L}_{int}^2 \ket{\small{\mbox{initial}}}$\footnote[1]{CC would like to thank Suvrat Raju for very useful discussions related to this.}. This term, $\mathfrak{L}_{int}^2$, will bring in the following combination of fields, 
 \be
 \sim \phi_1^2 + \phi_2^2 + 2 \phi_1 \phi_2~.
 \ee
We have mentioned earlier that, $\phi_1^2,\, \phi_{2}^{2}$ will correspond to the transitions to the 2-particle states of the respective scalar fields. In this article, we are specifically interested in the transition of the scalar fields from their respective vacua to 1-particle states, as for a moderate value of the acceleration of the detector, energetically 1-particle states are the most probable excited states for the quantum fields. Thus, for this choice of final particle states, we obtain a zero contribution from the terms like $\phi_1^2,\, \phi_{2}^{2}$, in the expression for transition amplitude. Therefore, the only novel term which will contribute in the transition amplitude, arises from $\sim \phi_1 \phi_2$. However, one can clearly observe that the vanishing contribution from the terms like $\phi_1^2,\, \phi_{2}^{2}$ and nonzero contribution from $\sim \phi_1 \phi_2$, is occurring due to the exclusive choice of the 1-particle states as the final particle states of the scalar fields. 

Generalisation of this multiple fields interaction to a set of $n$ number of scalar fields, now becomes straightforward. One can in principle generalise the eq.(\ref{scalar-2}) for the interaction of $n$ number of scalar fields and UD detector as following,
\be
\mathfrak{L}^{(n)}_{int}= \mathbbm{g^{(n)}}\, \mu(\t)(\phi_1[x(\t)]+ \phi_2[x(t)]+  \phi_3[x(t)]+.....+ \phi_n[x(t)])
\label{lagrangian_n}
\ee
However, in order to probe this novel structure at the level of the amplitude, with the consideration of simultaneous transition of all the scalar fields from their respective vacua to 1-particle states [as in fig.(\ref{states-2})], one  needs to go into the $n^{th}$ order in perturbation theory, where the only surviving term would be of the form :  $\sim \mu(\t) \phi_1[x(\t)] \phi_2[x(\t)] \cdots \phi_n[x(\t)]$. 
  
  Thus, in principle one can also acquire a distinct motivation for our chosen interaction Lagrangian in eq.(\ref{lagrangian_multiple}), as an effective description, originating from the $n^{th}$ order correction term in the perturbation series of the more fundamental Lagrangian as in eq.(\ref{lagrangian_n}), while one considers exclusively the 1-particle states to be the final states of all the quantum scalar fields. 
  
We urge to mention another crucial point to the readers that, this multiplicative structure of the Lagrangian ensures that in the final state of the system all the scalar fields must make a transition from their vacua to 1-particle states, in order to obtain a non-zero transition amplitude. One can immediately obtain from eq.(\ref{amp_multiple1}) that, if in the final state of the system, any one of the scalar fields remains in its corresponding vacuum, the transition amplitude will turn out to be zero.

From dimensional analysis one can see that in 4-dimensional spacetime, the coupling constant $\mathfrak{g}$ as in eq.(\ref{lagrangian_multiple}), acquires 
negative mass dimension for $n>1$. Thereby, our theory turns into a non-renormalisable theory. 
However, as explained in above discussion, this multiplicative Lagrangian can be viewed as an effective description, which has originated from the more fundamental interaction Lagrangian as in eq.(\ref{scalar-2}).  Eq.(\ref{scalar-2}), depicts the summation of the standard interaction Lagrangian of UD detector and single scalar field,  which is indeed a renormalisable theory. Therefore the non-renormalisability issue of the present model can be tackled in light of the effective field theory.

In this context, we urge to 
propose another possible fundamental renormalisable theory where one can consider 
an interaction Lagrangian of the following form:
\begin{eqnarray}
\mathcal{L}_{fund}=\sum_{i=1}^{n}\xi_1\phi_i\bar{\psi}\psi+\xi_2 \mu(\tau)\gamma^{\mu}A_{\mu}
+\xi_3\bar{\psi}A_{\mu}\gamma^{\mu}\psi \label{lagrangian_fund}
\end{eqnarray}
Here $\psi(x^{\mu})$, $A_{\mu}({x^{\mu}})$ is a fermion and gauge field respectively and $\gamma^{\mu}$ are the Dirac matrices. $\xi_1,~\xi_2,~\xi_3$ are dimensionless coupling constant. We conjecture that in 
the fundamental theory, the gauge field directly interacts with the UD detector at some higher energy scale and subsequently one 
can write down a low energy effective theory by integrating out the gauge field and fermion field contribution to obtain the detector and multiple scalar field coupling.
This conjecture can diagrammatically be shown:
\begin{figure}[H]
\includegraphics[scale=0.51]{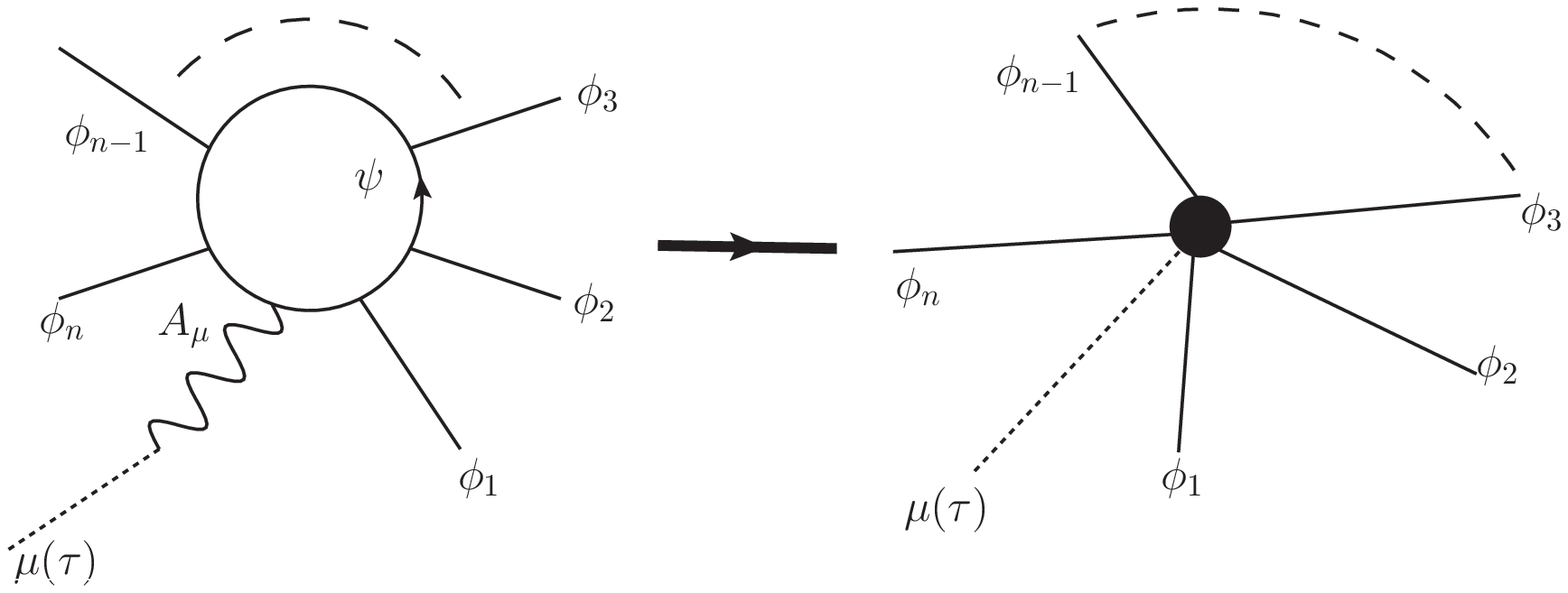}
\caption{Feynman diagram for the proposed fundamental theory and our chosen interaction Lagrangian}
\label{feynman}
\end{figure}
Although one must perform a more robust calculation describing the exact reduction to the low-energy effective theory, this description is given above provides a heuristic picture of the same. 
In this context, it is worth to be mentioned that, in \cite{kempf1}, the authors have proposed a renormalisation scheme to tackle the divergences originating due to the presence of non-linear coupling in the interaction Lagrangian. Subsequently, in some recent works \cite{Louko:2016ptn,Bradler:2017wdb}, such renormalisation schemes have been implemented in order to resolve the renormalisability issue of the respective theory. A more detailed inspection of the construction of a low energy effective theory (for our case) as derived from the fundamental theory and the choice of a proper renormalisation scheme is currently under progress. 

\section{Discussions and Outlook}\label{conclusion}
In the massless limit of the scalar fields, we observe that, for $n$ slightly greater than $1$, while the acceleration of the detector progress from $a<a_c$ to $a>a_c$, the change in the transition rate is not as drastic as compared with $n\gg 1$. The abrupt change in the transition rate with the variation of acceleration occurs for $n \gg 1$ and therefore the presence of critical acceleration becomes more apparent in the large $n$ condition. 

We interpret that, the UD detector starts detecting the quanta of radiation significantly when it simultaneously interacts with a large number of scalar fields and its acceleration crosses the threshold value, $a_c$.  As the temperature of the thermal bath is same as the acceleration up to a factor of $2\pi$, one can equivalently interpret our result as, in order to detect the quanta of these scalar fields the temperature of the thermal bath requires to cross a certain temperature, called critical temperature ($T_c$). 

Physically, we describe that the outcomes of the present work are based on the nature of the chosen interaction, dependence of the transition rate on the number of scalar fields and the presence of critical acceleration. The multiplicative structure of the interaction Lagrangian dictates that in order to get a non-zero transition amplitude, all the scalar fields must make a simultaneous transition from their respective vacua to 1-particle states. If a small number of scalar fields are interacting with the detector, the response function of the detector is not large enough in comparison to the standard UD detector case, even under the condition $a>a_c$. Whereas, upon applying both the conditions, $n\gg 1$, and $a>a_c$, the response function of the detector shows a drastic change than the standard UD case. 
 Therefore, the enhancement of the transition rate is occurring due to the collective interaction of a large number of fields and UD detector, while the detector crosses a threshold acceleration $a_c$.

Our analysis is strongly implying that the appearance of critical acceleration, enhanced response function of the detector, are sole outcome of the UD detector and {\it multiple} scalar fields interaction, which cannot be perceived in UD detector and single field scenario. 

We comment about some of possible future works and extensions of this theory such as it would be interesting to examine the UD detector and multiple gauge fields, fermion fields interaction by following the present analysis. Observing the interesting features for the massless limit of the scalar fields, we are intending to study the $\f{m}{a}\ll1$ limit of the theory in a follow up work. 
There is some liberty in the multiple fields and detector interaction scenario over the standard case such as one can study the transition rate by relaxing the constraint of simultaneous interaction or by choosing some different set of final states of the system. These studies may also lead to many interesting behaviours of quantum fields and provide a further understanding of the respective quantum theories, from the perspective of an accelerated frame of reference.
Moreover, one can in principle examine that how the interaction of the multiple quantum fields and UD detector may put some constraints on the theories dealing with large number of species as described in \cite{Dvali:2007hz,Dvali:2009ne,Dvali:2009fw}.
\vskip 1.5 mm
\noindent
{\bf Acknowledgement:} We thank Debashish Borah, Girsh S. Setlur, Krishnendu Sengupta, Loganayagam R, Soumitra Sengupta, Tanmoy Paul for several useful discussions during different stages of the project. We especially thank Suvrat Raju for his insightful comments. We would like to express our heartfelt gratitude to Stephen A. Fulling for his invaluable suggestions and inspiring comments regarding our work. We would also like to thank to the anonymous referee for his/her valuable comments.  
\appendix 
\section{Analysis of Bessel Function}\label{Analysis of Bessel Function}
We use the following properties of Bessel function :
\begin{eqn}
I_{\a}(x) &= i^{-\a} J_{\a}(ix) \\
K_{\a}(x) &= \f{\pi}{2 \sin(\a \pi)} \ \bigg[ I_{-\a}(x) - I_{\a}(x) \bigg]
\end{eqn}
and obtain : $K_{\a}(x) = \f{\pi}{ 2 \sin(\a \pi)} \ \bigg[ i^{\a} J_{-\a}(i x)  - i^{- \a} J_{\a}(i x)  \bigg]$\\
This satisfies an important recursion relation, 
\be
K_{\nu - 1}(z) + K_{\nu + 1}(z) = - 2 \ \f{d}{dz} K_{\nu}(z)
\ee
In our analysis, we expand the Bessel function under the two conditions such as small and large argument which correspond to the small and large mass limit for our case:
\begin{eqn}
K_{\nu}(z) \approx \begin{cases}
&z \rarrow 0 : \ 2^{-1 - \nu} z^{\nu} \ \G(-v) + 2^{-1 + \nu} z^{- \nu} \ \G(\nu) \\ \\
&z \rarrow \infty : e^{-z} \sqrt{\pi/(2 z)} 
\end{cases}
\end{eqn}
For $\nu = 1$, $K_{\nu}(z \rarrow 0)$ takes a simpler form, 
\begin{eqn}
\lim_{z \rarrow 0} K_1(z)  = \f{1}{z} + \f{z}{4} \bigg[ 2 \g - 1 - \log{4} + 2 \log{z}  \bigg]
\end{eqn}
\section{The ``Large $n$'' Limit} \label{large_n}
We study the large $n$ limit of the transition rate by using Stirling's Approximation. The dominating term in $F_n(\o)$ in eq.(\ref{large n power}), comes from the function $I(\o, a, n)$ which is depicted in eq.(\ref{Integral I}). This can be seen as follows, 
\begin{eqn}
F_{n \gg 1}(\o) \approx \mathfrak{g}^2 \l(\f{a}{2\pi}\r)^{2n}  \f{1}{(2n - 1)!} \prod_{k = 1}^{n} \bigg[ \f{\o^2}{a^2} + (n - k)^2 \bigg]
\label{appendix_1}
\end{eqn}
This can be further simplified as, 
\begin{eqn}
F_{n \gg 1}(\o) \approx \mathfrak{g}^2 \l(\f{a}{2\pi}\r)^{2n}  \f{(n - 1)!^2}{(2n - 1)!} \bX{n-1}{\f{\o}{a}}\label{appendix_2}
\end{eqn}
Here the function $\bX{n}{\o/a}$ is a modification of the Pocchammer function which takes the following form, 
\be
\bX{n}{\b} = \l( \b^2 \r) \l(1 + \f{\b^2}{1^2} \r) \l(1 + \f{\b^2}{2^2} \r) \cdots \l( 1 + \f{\b^2}{n^2 }\r)
\label{pocchammer}
\ee
It is clear that the asymptotic behaviour of the function $F_n(\o)$ is decided by the limit $n \rarrow \infty$. In this analysis for $n\rightarrow \infty$, one needs to consider with a little more caution is  $\o \rarrow \infty$. In this case, we have the factor of $\big( e^{2\pi \o/a} -1  \big)^{-1}$ which diverges faster than $\bX{n}{\o}$ for $\o \rarrow \infty$. Thus for the case $\o \gg 1$, one can expect that the transition rate ceases to zero. 

Therefore we concentrate on the limit $n \rarrow \infty$ for a finite value of $\o$. In this case, we use the Stirling's approximation to the ratio of factorials in eq.(\ref{appendix_2}). 
\begin{proof}
Stirling's Approximation, for $n \gg 1$
\be
n! \approx \sqrt{2 \pi n} \l(  \f{n}{e}  \r)^n \label{stirlingapprox1}
\ee
In the limit $n \gg 1$, we have,
\be
\f{(n-1)!^2}{(2n-1)!} \approx \f{n}{n^{1/2}} \f{(n/e)^{2n}}{(2n/e)^{2n}} =  \f{1}{2^{2n}} \ \f{1}{\sqrt{n}}
\ee
\end{proof}
Hence, $F_{n}(\o)$ in the limit $n \gg 1$ becomes, 
\be
F_{n\gg 1}(\o) \approx \f{1}{\sqrt{n}} \l( \f{a}{a_c} \r)^{2n}
\ee
Where $a_c = 4\pi \a$. Thus we see that an acceleration scale appears naturally in the asymptotic limit of $F_{n}(\o)$. For $a > a_c$, one sees that the transition rate diverges for large $n$ as dictated by the term $( \f{a}{a_c})^{2n}$, and for $a < a_c$, the rate converges. 

\end{document}